\documentclass[useAMS,usenatbib]{mn2e}
\usepackage{graphicx}
\def\p1{\phantom{1}}

\def\simless{\mathbin{\lower 3pt\hbox
     {$\rlap{\raise 5pt\hbox{$\char'074$}}\mathchar"7218$}}}   %< or of order
\def\simmore{\mathbin{\lower 3pt\hbox
     {$\rlap{\raise 5pt\hbox{$\char'076$}}\mathchar"7218$}}}   %> or of order
\def\hide#1{}

\title[On the kHz QPO frequency correlations]
{On the kHz QPO frequency correlations in bright neutron star X-ray binaries}

\author[T. Belloni, M M\'endez and J. Homan]{T. Belloni$^{1}$\thanks{E-mail:
tomaso.belloni@brera.inaf.it},
M. M\'endez$^{2}$\thanks{E-mail: mariano@sron.nl},
J. Homan$^{3}$\thanks{E-mail: jeroen@space.mit.edu}\\ 
$^{1}$INAF-Osservatorio Astronomico di Brera, Via E.
Bianchi 46, I-23807 Merate (LC), Italy\\ 
$^{2}$SRON, Netherlands Institute for Space Research, Sorbonnelaan 2,
3584 CA Utrecht, The Netherlands\\ 
Astronomical Institute Anton Pannekoek, University of Amsterdam,
Kruislaan 403, 1098 SJ Amsterdam, The Netherlands\\
$^{3}$MIT Kavli Institute for Astrophysics and Space Research, 70 Vassar Street,
Cambridge, MA 02139, USA }

\begin{document}

\date{
Accepted 2007 January 10. Received 2007 January 10; in original form 2006 November 24}

\pagerange{\pageref{firstpage}--\pageref{lastpage}} \pubyear{2007}

\maketitle

\label{firstpage}

\begin{abstract} We re-examine the correlation between the frequencies of upper and
lower kHz quasi-periodic oscillations (QPO) in bright neutron-star low-mass
X-ray binaries.  By including the kHz QPO frequencies of the
X-ray binary Cir X-1 and two accreting millisecond pulsars in our
sample, we show that the full sample does not support the class of
theoretical models based on a single resonance, while models based on
relativistic precession or Alfv\'en waves describe the data better.
Moreover, we show that the fact that all sources follow roughly the same
correlation over a finite frequency range creates a correlation
between the linear parameters of the fits to any sub-sample.

\end{abstract}


\section{Introduction}

Thanks to the Rossi X-Ray Timing Explorer ({\em RXTE}), the past
decade  has seen the discovery of high-frequency quasi-periodic
oscillations (QPOs) in neutron-star X-ray binaries \citep[see][for a
review]{van-der-Klis-05}. These oscillations often appear in pairs,
with frequencies ranging from a few hundred Hz to more than 1 kHz,
hence the name {\em kHz QPOs}. Such fast variability provides a probe
into the accretion flow very close to the compact objects and is a
powerful tool to observe effects of general relativity. The
frequencies of the kHz QPOs are strongly correlated with other timing
and spectral features \citep*[see][]{Ford-98, Psaltis-99, Belloni-02,
van-Straaten-04}.

Several models have been proposed for these oscillations; these
models are based on the identification of the QPO frequencies with
various characteristic frequencies in the inner accretion flow
\citep{Stella-99, Osherovich-99, Zhang-04, Lamb-03}. However, there
is still no consensus as to the origin of the kHz QPOs. 

QPOs in the range $30-450$ Hz have been observed in a limited number
of black-hole candidates; in some of these systems the QPOs seem to
move in frequency \citep[see e.g.][]{Homan-01}. However, in the four
cases when two QPOs are detected simultaneously in a given black-hole
candidate, these QPOs appear at frequencies that are consistent with
being always the same, with frequency ratios that are consistent
with the ratio of two integer numbers, such as 3:2 and 5:3 \citep[see
e.g.][]{Strohmayer-01a, Strohmayer-01b, Miller-01}. A model for these QPOs 
based on the resonance of characteristic frequencies at a specific radius in
the accretion disc was presented by \cite{Abramowicz-01}.

\cite{Abramowicz-03} reported that the ratios of the frequencies in the
kHz QPOs from the brightest low-mass X-ray binary (LMXB) in the sky, Sco
X-1, tend to cluster around a value of 1.5, which they interpret as
evidence for a 3:2 resonance also being responsible for the QPO pairs
in neutron-star systems. 

\citet*[][hereafter BMH05]{Belloni-05}, have shown that the
frequencies of the lower, $\nu_{low}$, and upper, $\nu_{high}$, kHz
QPOs are strongly correlated along a line which is inconsistent with
$\nu_{high} = 1.5 \nu_{low}$, but can be roughly approximated as a
straight line not passing through the origin. They also showed that
when this correlation is taken into account, the peak in the
distribution of ratios in the Sco X-1 data used by
\cite{Abramowicz-03} is only marginally significant, since the
distribution of the ratio of two correlated quantities is completely
determined by the distribution of one of them. BMH05 also found that
the frequency distributions of lower kHz QPOs for five systems have
statistically significant peaks (some of which did not occur near a
3:2 frequency ratio). However, those peaks are consistent with being
caused by a (nearly) random walk of the QPO frequency with time,
together with the sampling of the observations.

A mathematical approach to the resonance
model was presented by \cite{Rebusco-04} in order to explain the Sco X-1
results; \cite{Rebusco-04} attributed the discrepancies between the data
and a pure 3:2 ratio to the action of an additional {\em ad-hoc} force
which moves the frequencies away from their `natural'
ratio, along a correlation such as the observed one.

While the correlation analyzed in BMH05 treated the sample of sources
as a whole, \citet[][hereafter Z06]{Zhang-06} analyzed the sources
separately and showed that there are differences between sources, and
deviations from a linear relation.
\citet[][hereafter A05a and A05b]{Abramowicz-05, Abramowicz-06} presented
an analysis of the same correlations. Fitting a number of sources
separately with a linear function $\nu_{high} = a\nu_{low} +  b$, they
obtain a set of $a,b$ pairs (one pair per source). The pairs $a$,$b$
themselves follow a linear relation $b=Aa+B$ with intercept $B$ close to
1.5. A05a and A05b interpret this fact as a property in accordance to the
resonance model of \cite{Abramowicz-01}.

Here we show that the analysis of separate sources presented by A05a
and A05b does not add new information to that of the global
correlation. We also re-examine the frequency correlation with the
addition of new data from the literature. In particular, the newly
discovered high-frequency oscillations in Cir X-1
\citep{Boutloukos-06}, which extend the correlation to lower
frequencies, are important to distinguish between the different
classes of models. We show that the frequencies of the kHz QPOs in 
Cir X-1 are inconsistent with the resonance model of
\cite{Abramowicz-01}.

\section{Additional data}

Since the analysis by BMH05, new pairs of kHz QPOs frequencies 
have been published, some of which
are particularly important for checking and extending the
correlation between the kHz QPO frequencies. 

\subsection{Cir X-1}

\begin{figure*}
\begin{center}
\includegraphics[angle=0,width=16.cm]{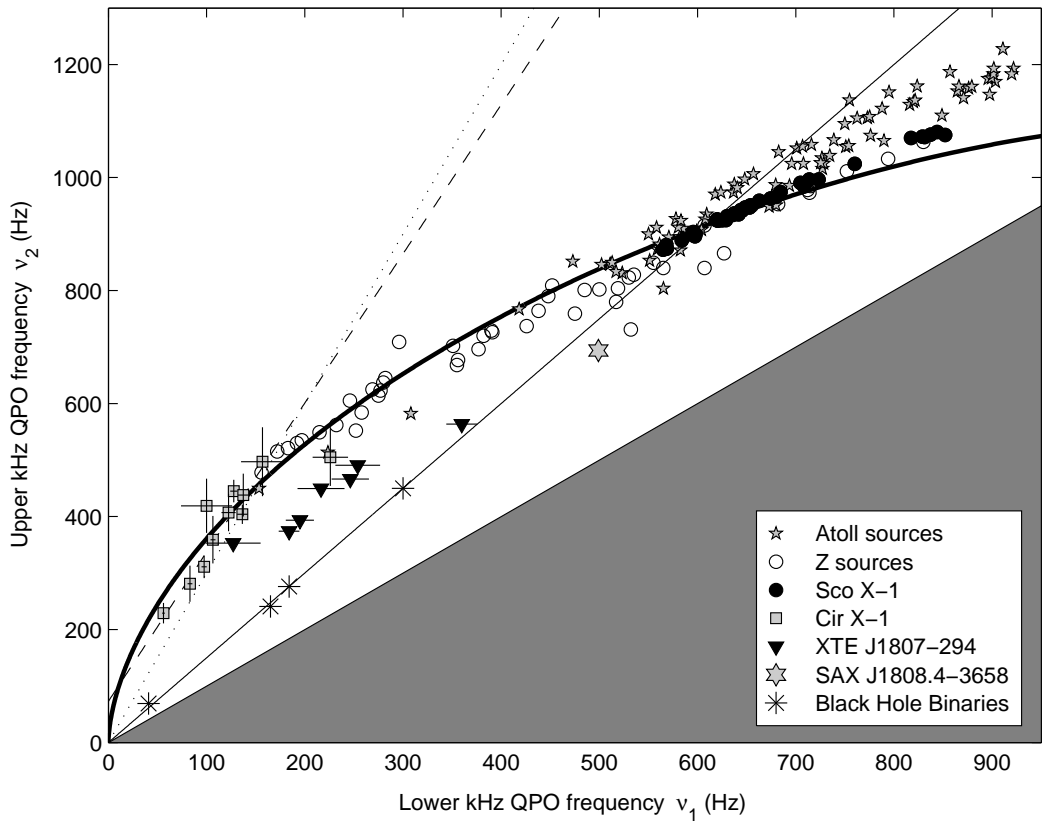}
\end{center}

\caption{Plot of the kHz QPO frequency-frequency relation including
the points from Cir X-1 \citep[]{Boutloukos-06}. The XTE J1807--294
data are from \citet{Linares-05} and the  SAX J1808.4--3658 point is
from \citet{Wijnands-03}. The asterisks
are pairs of confirmed simultaneous high-frequency QPOs in black-hole
systems. The thin solid line represents a fixed 3:2 ratio, and the
dotted line a 3:1 ratio. The dashed line is a fit to the Cir X-1
points, excluding the highest frequency one. The thick solid line
shows the relation between the periastron-precession and the
Keplerian frequencies for $M = 2 M_\odot$; in the relativistic
precession model \citep{Stella-99} those two frequencies have been
identified with the lower and upper kHz QPO, respectively. The dark
area represents $\nu_{low} > \nu_{high}$, obviously not allowed by
definition.}

\label{figure_correlation}
\end{figure*}

Recently, twin kHz QPOs have been discovered in the peculiar LMXB Cir
X-1 \citep{Boutloukos-06}. At variance with what is observed in Z and
atoll sources (see e.g., BMH05), the kHz QPO frequencies observed in this 
system are substantially lower than in the Z and atoll sources 
(in the range 56--225 Hz and
230--500 Hz for the lower and upper kHz QPO, respectively) and
their difference {\it increases} with frequency.  Adding the 11 pairs of
kHz QPO frequencies from \cite{Boutloukos-06} to the
frequency-frequency correlation of BMH05, we find that,
compared to the other kHz QPO sources, they follow a
very different correlation (see Figure \ref{figure_correlation}), with
linear best-fitting parameters $a=2.64 \pm$0.26 and $b=73 \pm$28
(excluding the highest frequency point, which clearly would tilt a
linear relation). We note that this relation intersects the $3/2$
line at $\nu_1 = - 64 \pm$ 29 Hz. \cite{Boutloukos-06} mention that
for Cir X-1 a constant $\nu_{low}$:$\nu_{upp}=$ 1:3 ratio is
possible. A linear relation with $a=3$ and $b=0$ (dotted line in Figure
\ref{figure_correlation}) fits the data  statistically worse than the $b \neq 0$
one (reduced $\chi^2$ of 2.4 and 1.3 respectively);  a fit with $b=0$
and $a$ free, with a reduced $\chi^2$ of 1.3, yields $a=3.31\pm
0.10$, i.e. inconsistent with a value of 3.

%-----------------------------------------------------------------------
%-----------------------------------------------------------------------

\subsection{XTE J1807--294}

Another important source to include in the sample is the accreting
millisecond pulsar XTE J1807--294, for which pairs of kHz QPOs have
been discovered by \cite{Linares-05}. It should be noted that
some of the peaks found in this source are rather broad.  However,
in seven cases out of eight, \cite{Linares-05} could firmly identify two
peaks as lower- and upper-kHz QPOs based on
correlations with low-frequency features. We consider only these
seven pairs as {\it bona fide} kHz QPOs.
These points are included in the samples analyzed by A05b. 
Similar values were reported by \cite{Zhang-06b} using the same data.

\subsection{SAX J1808.4--3658}

For this system, the first discovered accreting millisecond pulsar,
\cite{Wijnands-03} found a pair of kHz QPOs at $\sim 500$ Hz and
$\sim 695$ Hz, respectively. In addition, \cite{van-Straaten-04}
reported pairs of broad features at relatively low frequencies that
they interpreted as the lower kHz QPO in this system; however, they
also mention the possibility that these QPOs are a different
component altogether. We do not include these points in our
sample.  The point corresponding to SAX J1808.4--3658 is
shown in Figure \ref{figure_correlation} as a six-pointed star.

\section{Properties of the global correlation vs. single-source
correlations}

In the following we discuss the significance of correlations between 
linear parameters derived from separate sources. We first show the results
of simulations based on the existing data. This more intuitive approach is then
followed by its analytical interpretation.

\subsection{The BMH05  sample and single sources}

As a starting point, we take  the sample originally published by
BMH05 and partially used also by Z06, A05a and A05b, plus the
recent frequencies from the millisecond accreting X-ray pulsar XTE
J1807--294 \citep[see above]{Linares-05}. Different classes of sources
(Atoll and Z) can follow different correlations, as shown by BMH05,
A05a, and A05b, and the same applies to individual sources (see also
Z06). However, the fact that the overall BMH05 sample of points
shows a good degree of correlation (we will assume for the remainder of
the paper that the correlation is linear, but the precise functional
shape is not relevant for this discussion), indicates that the
differences between source classes and individual sources cannot be
very large. 

In Figure \ref{figure_single}, we plot the BMH05 sample together with the
best-fit  models found in A05a and A05b for selected sources. Notice
that discrepant $a$ and $b$ values are quoted in A05a and A05b for
the same sources; for the discrepant cases, we use the A05a values.
We remark that Figure \ref{figure_single} contains also points from sources not
fitted with the model lines plotted in that figure. Unless there is
an unlikely correlation between dynamic range in frequencies and
deviation from the average correlation, the moderate spread observed
in the data excludes large deviations in correlation parameters for
single sources. In other words, in Fig.  \ref{figure_single} the spread of the
single-source correlations, when extrapolated over the full range of the sample,
is larger than the spread of the points. Slopes very different from 
that of the sample correspond to sources for which a reduced dynamic
range of frequencies is observed.
This leads to the conclusion that it is likely that
these different correlations are {\it the result}, at least in part, of the limited
range of kHz QPO frequencies  covered by these sources.
Notice that, at variance with what reported by A05a, the best fit models to 
different sources do not intersect at the same point.

\subsection{The sample as a whole}

\begin{figure}
\begin{center}
\includegraphics[angle=0,width=8cm]{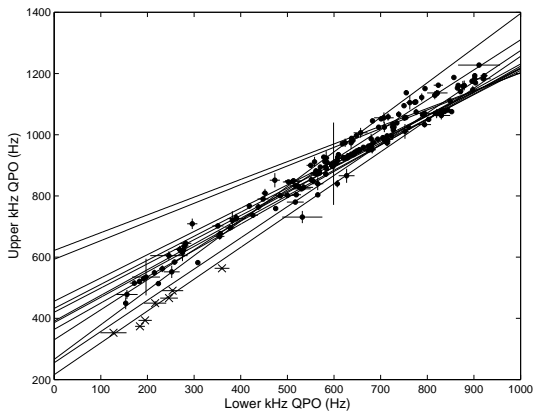}
\end{center}
\caption{Frequency-frequency correlation for the sources in BMH05, with
the addition of the centroid frequencies of XTE J1807--294 from Linares
et al. (2005) [crosses, see text]. The lines are the best fit lines
from A05a and A05b.}
\label{figure_single}
\end{figure}

As shown by BMH05, the lower and upper kHz QPO frequencies of a large
sample of sources are roughly linearly correlated 
as $\nu_{upp}=0.92 \nu_{low} + 360$. The precise
values are irrelevant here, as well as the precise functional shape
of the correlation (see also Z06). We consider here the BMH05 sample
{\em without} taking into account the association of points with separate
sources. We simulate artificial sources by taking 1000 random
sub-samples of $n$ real points where $n$ is also a random number for
each artificial source, chosen between 20 and 80. For each
sub-sample, we fit the $n$ values with a linear function and thus
derive a set of 1000 pairs of $a$ and $b$. A plot of these pairs
shows a marked anti-correlation $b=Aa+B$ with slope
$A=(-1.69\pm 0.01)\times 10^{-3}$ and intercept $B=1.51
\pm$0.01, consistent with the values quoted by A05b from their source-by-source
analysis (see Figure \ref{figure_simul}; the range in $a$ and $b$ values in
our simulation is limited because the procedure does not likely
produce random subsamples spanning a limited range in frequency, as
it appears to happen in reality.) From this simulation we find that
any randomly selected subset of points from this dataset yields $a,b$
pairs that are distributed along such a straight line, and have
$B\sim1.5$. Simulating random straight lines in the plane would of
course not lead to any $a$ vs. $b$ correlation, but in this case we
are extracting lines that cross the correlation segment visible in
Figure \ref{figure_single}.

%-----------------------------------------------------------------------
%-----------------------------------------------------------------------

The origin of the correlation shown in Figure \ref{figure_simul} can be seen
more clearly  through a simulation that does not make use of the
actual observed frequencies. We simulated an artificial set of pairs
of frequencies following a linear correlation $y=0.92x+360$ 
over the range $\nu_1$=150--900 Hz and added
a Gaussian spread with $\sigma$=50 Hz. Then, we repeated 100 times
the procedure outlined above (extract 1000 random subsamples, fit
them with a linear function, fit linearly the $a$ values vs. the $b$
values and obtaining the intercept $B$), obtaining a set of 100
values for $B$. 
The resulting distribution of the 100 $B$ values  is
strongly peaked around 1.5; this shows that, while this value could originate
from the differences in the frequency-frequency distributions for different sources,
as suggested by A05a,b, it could also simply be the consequence of the values
$a$=0.92 and $b$=360 of the global correlation {\em and} of the ranges
spanned by the observed frequencies.
We repeated the simulation with the same $a$ and $b$ values, but 
with $\nu_1$ in the ranges
600--1000 Hz and 10-500 Hz, and obtained average $B$ values of
1.36 and 2.0 respectively.

\subsection{Significance of the anti-correlation between linear parameters}

The results of the simulations shown above can be interpreted analytically.
The $a$,$b$ parameters of a $y=ax+b$ correlation are covariant. This means that
the confidence contours around the best fit value are always elongated diagonally
in the direction of anti-correlation. The inclination of the confidence ellipse is 
determined both by the best fit parameters and the distribution of the points along
the correlation. In particular, the intercept $A$ of the major axis of the ellipse can be
found simply by fitting a linear relation of the form $y=Ax$ to the data pairs $x_j,y_j$.
Analytically, this can be calculated as
\begin{equation}
A = \sum_j {{y_j x_j\over \sigma_j}} /  \sum_j {{x_j^2\over \sigma_j}}
\end{equation}
If the errors are all the same size, Eqn. 1 reduces to 
\begin{equation}
A = <y x > / <x^2>
\end{equation}
For the typical range of $x=\nu_1$ and $y=\nu_2$ in the data, this can be
approximated as $A=<\nu_2> / <\nu_1>$
Obviously, a fit with zero intercept to data linearly distributed with non-zero intercept
will cross the data around the middle of the points, which is roughly centered 
at $<\nu_2> / <\nu_1>$. This means that,
given our datasets, the values of $A$ will depend also crucially on the distribution
of points (see also BMH05). 

We selected a clean sample of frequencies from a few sources, fitted with a linear
function with zero intercept and computed the A values from Eqn. 1, which resulted
compatible with the fitting parameters
(see Table \ref{table1}, where $A$ values obtained with Eqn. 1 are shown, as well as
$<\nu_2>/<\nu_1> $ values). Notice that for GX 17+2 and
GX 5--1 we only selected pairs of frequencies where both detections were above
3$\sigma$ in Homan et al. (2002) and Jonker et al. (2002a). For GX 340+0, this
selection from the data in Jonker et al. (2000) resulted in only three pairs of 
frequencies: we therefore excluded this source from the sample.
It is clear that the $A$ (excluding Cir X--1) values cluster around the value 1.5. 
Fig. \ref{figure_ellipses} shows the best-fitting A and B values and confidence contours for these sources. The alignment of all ellipses is evident.

\begin{table*}
\begin{tabular}{lccccc}
\hline
Name& $a$ & $b$ & $A$ &$<\nu_2>/<\nu_1> $&Ref.\\
\hline
GX 5--1      &   0.856 $\pm$ 0.022  & 374 $\pm$ 8   &1.85&1.95&1\\
Sco X--1     &   0.791 $\pm$ 0.005  & 430 $\pm$ 3   &1.41&1.42&2\\
GX 17+2     &   0.870 $\pm$ 0.041  & 361 $\pm$ 26 &1.45&1.46&3\\
4U 0614+09&  1.078 $\pm$ 0.020  & 269 $\pm$ 12 &1.49&1.52&4,5\\
4U 1728--34&  0.899 $\pm$ 0.036  & 418 $\pm$ 25 &1.47&1.46&6\\
4U 1636--53&  0.672 $\pm$ 0.025  & 541 $\pm$ 22 &1.32&1.34&7,8\\
4U 1608--52&  0.810 $\pm$ 0.027  & 419 $\pm$ 16 &1.45&1.42&6\\
4U 1820--30&  0.892 $\pm$ 0.045  & 355 $\pm$ 33 &1.39&1.39&9\\
4U 1915--05&  1.157 $\pm$ 0.045  & 256 $\pm$ 20 &1.59&1.62&10\\
XTE J1807--294&  1.145 $\pm$ 0.124  & 171 $\pm$ 30 &1.90&1.95&11\\
\hline
Cir X--1      &  2.339 $\pm$ 0.474  & 104 $\pm$ 58&3.11&3.18&12\\
\hline
\end{tabular}
\caption{Sample fo LMXBs with best fit linear parameters, $A$ values
obtained with Eqn. 1,  $<\nu_2>/<\nu_1> $, and 
references. 1. Jonker et al. (2002a); 2. M\'endez \& van der Klis (2000); 3. Homan et al. 
(2002); 4. van Straaten et al. (2000); 5. van Straaten et al.
(2002); 6. M\'endez et al. (2001); 7. Di Salvo et al. (2003); 8. Jonker et al. (2002b); 
9. Zhang et al. (1998); 10. Boirin et al. (2000); 11. Linares et al. (2005);
12. Boutloukos et al. (2006).
}
\label{table1}
\end{table*}

As we have shown above, the values of $A$ cluster around the value 1.5 because
the observed linearly-correlated frequencies show a ratio of mean values close 
to 1.5. This was also shown in BMH05. This means that the confidence contours will
on average have an inclination pointing at $B=0$,$A=1.5$. 

\section{Discussion}

Aside from precise fitting of separate sources with different models,
it is evident from Figure \ref{figure_single} that the centroid frequencies
of kHz QPOs of both atoll and Z sources are roughly distributed,
although with some scatter, around a line approximated by $\nu_{high}
= 0.92 \nu_{low} + 360$ (see BMH05). 
We have shown that this fact
alone leads to a strong anti-correlation, with intercept very close
to the value $1.5$ (the exact value of the intercept depends on the 
range of frequencies spanned by the QPOs),
between the parameters $a$ and $b$ obtained from
linear fits to any sub-sample of points, or to points generated
randomly along the global correlation. This result alone implies that
the analysis of separate sources cannot be used 
in the way proposed by A05a,b for testing the resonance model.

%------------------------------------------------------------------------------------------------
\begin{figure}
\begin{center}
\includegraphics[angle=0,width=9cm]{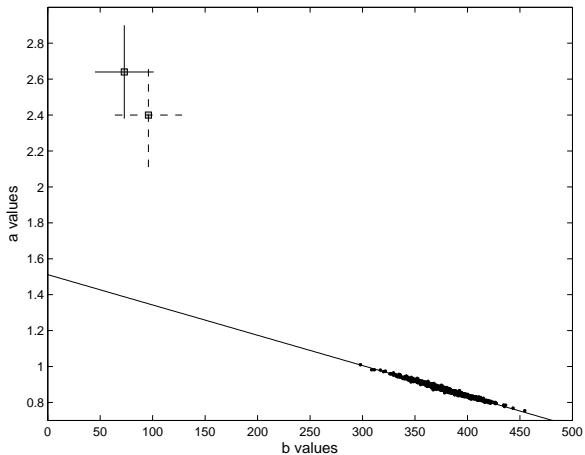}
\end{center}
\caption{Linear parameters $a$ and $b$ obtained from 1000 random
samples from the points in Figure \ref{figure_single}. The points lie on a
correlation which is itself linear with intercept $B$=1.51$\pm$0.01.
The squares correspond to the $a$,$b$ pair obtained for Cir X-1
including the highest-frequency point (full error bars) and without
(dashed error bars).}
\label{figure_simul}
\end{figure}
%------------------------------------------------------------------------------------------------

The question however remains why the frequencies are distributed
along that major correlation and with these frequency boundaries.
As shown above, the intercept of $1.5$ in the $a$-$b$ anti-correlation could
be different should the linear parameters of the sample of sources,
or the range spanned by the kHz QPO frequencies be different. 
Moreover, notice that there are different combinations of $a$,$b$ and frequency range
that yield to $B\sim$1.5.
The global correlation visible in Figure  \ref{figure_correlation}, and its frequency 
range, constitute therefore one of the major observational results that must be
directly addressed by theoretical models. 

Figure \ref{figure_correlation} shows that the frequency points from
Cir X-1 pose a more serious problem to the resonance
model of \cite{Abramowicz-03}: They are distributed along a steeper
line than the one followed by the frequencies of the sources in
BMH05, A05a, and A05b, constituting a low-frequency extension of the
correlation defined by those sources. Notice that the linear
parameters $a$ and $b$ for Cir X-1 are very different from
those of the other sources, in a way that is {\it not} at all
consistent with the correlation shown in Figure 2 of A05b (see Figure \ref{figure_simul}).

%-----------------------------------------------------------------------
\begin{figure}
\begin{center}
\includegraphics[angle=0,width=8cm]{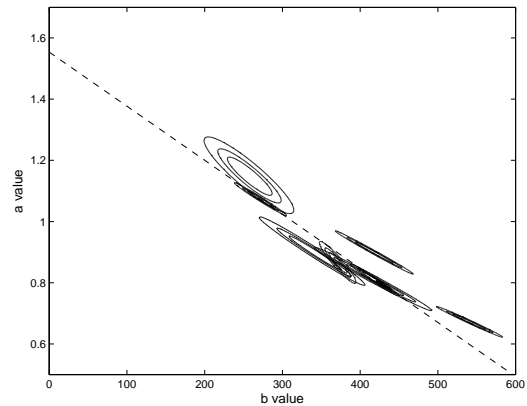}
\end{center}
\caption{Confidence contours (68.3\%, 90\%, 99\%) corresponding to the linear
fits to the sources in Table \ref{table1}, with the exception of XTE J1807--294 
and Cir X--1. The dashed line is the linear fit to the data, yielding a slope of 
1.55$\pm$0.07 and an intercept of -0.0018$\pm$0.0002.
}
\label{figure_ellipses}
\end{figure}
%-----------------------------------------------------------------------

It is clear from Figure \ref{figure_correlation} that the frequencies of Z and
atoll sources do not follow a relation $\nu_{high}=1.5\nu_{low}$, as
already pointed out by BMH05. \cite{Rebusco-04} presented a model
where the addition of an {\em ad hoc} force on an oscillator responsible for an
otherwise fixed pair of frequencies could lead to the observed
correlation. The kHz QPOs in Cir X-1 \citep{Boutloukos-06} follow a
correlation which intersects the 3:2 fixed-ratio line at $\nu\sim
-64$ Hz, i.e. the two lines do not intersect in the physical range of
interest for the frequencies. This, together with the fact that the
kHz QPOs in Cir X-1 show other properties and correlations very
similar to the kHz QPOs in other neutron-star systems, cast doubt on
the model of \cite{Rebusco-04}. 
The possibility of
Cir X-1 being a black hole is very unlikely. Eight X-ray bursts were
observed with EXOSAT \citep*{Tennant-86,Tennant-86b}, the
high-frequency QPOs are consistent with being neutron-star kHz QPOs, and
at high luminosities the tracks in the X-ray color-color diagrams and the 
low-frequency variability are very similar to those for the Z sources
\citep{Shirey98, Shirey99}. Also, it is suggestive
that the Cir X-1 frequencies, while following a much steeper linear
relation, join smoothly with the low-frequency end of the Z sources.

One of the most interesting results from \cite{Boutloukos-06} is that
the Cir X-1 points are the first kHz QPOs observed with a frequency
difference which {\it increases} with frequency, in accordance with
the prediction of the relativistic precession model \citep{Stella-99}
and the Alfv\'en wave model \citep{Zhang-04, Zhang-05, Zhang-06}. 
In Figure \ref{figure_correlation}, we also plot the line
for the relativistic precession model corresponding to a mass of
2$M_\odot$ (the line for the Alfv\'en wave model lies very close to
that of the relativistic precession model). The agreement with Cir
X-1 and the Z sources at low frequencies is remarkable.  At high
frequencies, the points lie above the model line; this of course
could reflect different masses for the neutron stars, but for some
sources additional assumptions are required anyway
\citep[see][]{Stella-99}.  Moreover, above $\nu_1$=150 Hz the data
follow a rather straight correlation, while the model line is
significantly curved at all frequencies.

The points corresponding to the two accreting millisecond pulsars,
XTE J1807--294 and SAX J1808.4--3658 \citep{Linares-05, Wijnands-03}
lie below the global correlation. It is interesting
to note that from their analysis of correlations between all timing
frequencies of a sample of atoll sources and accreting millisecond
pulsars (not including XTE J1807--294), \cite{van-Straaten-05}
conclude that for two of the millisecond pulsars (including SAX
J1808.4--3658) both the lower and upper kHz QPOs appear shifted down
by a factor $\sim 1.5$ \citep[][have to multiply $\nu_{low}$ and
$\nu_{upp}$ of these two millisecond pulsars by $\sim 1.5$ for them
to match the correlation of the other sources in their
sample]{van-Straaten-05}. Applying a 1.5 factor to both kHz QPO
frequencies corresponding to the pulsars in our sample, the two
sources are indeed moved to the global correlation.
Within the relativistic precession model, such a shift could imply a
neutron-star mass higher by a factor of 1.5, which is unlikely as it
would lead to accreting millisecond pulsars containing neutron stars
of 3$M_\odot$, assuming 2M$_\odot$ for the other sources.  
We note that a similar shift does not move Cir X-1
into the linear correlation defined by the other sources.

On  Figure \ref{figure_correlation}, we also plot the four points corresponding
to the pairs of confirmed simultaneous high-frequency QPOs from
black-hole systems (GRS 1915+105: Strohmayer 2001a; GRO J1655--40:
Strohmayer 2001b; XTE J1550--564: Miller et al. 2001; H 1743--322:
Homan et al. 2005). Three of these points lie very close to the 3:2
line, while GRS 1915+105 is consistent with a 5:3 ratio (it appears
close to the 3:2 line due to the low frequencies involved). 
Clearly, for black hole binaries, a model in which the QPO frequency ratio
is the ratio of two integer numbers is consistent with the available data.

\section{Conclusions}

We showed that the correlation between parameters obtained from 
linear fitting to separate sources
applied to atoll and Z sources by A05a and A05b is a direct result of
the observed general correlation and cannot be used for testing their
models. Moreover, the newly discovered kHz QPOs in the peculiar
neutron-star binary Cir X-1 follow a different correlation than atoll
and Z sources. This observed correlation appears to be in contradiction with
the predictions of all modifications of resonance models, but is in
good general agreement with the relativistic precession model
(although it is known that the model in its simplest version does not
explain all observables, such as the relation between kHz QPOs and
spin frequency, and needs correction at high frequencies) and the
Alfv\'en wave model. Our results indicate that the observed
high frequency QPOs in black hole and neutron star low-mass X-ray binaries 
most likely have a different origin.

\section*{Acknowledgments}

TB acknowledges financial contribution from contract ASI-INAF I/023/05/0.
The Netherlands Institute for Space Research (SRON) is supported 
financially by NWO, the Netherlands Organization for Scientific Research.
We are grateful to an anonymous referee for a careful
reading of the manuscript and for comments that helped us improve it,
especially on the issue of the covariance of the parameters of the
indivdual fits.

\label{lastpage}

\end{document}